\documentclass[12pt]{article}
\usepackage{a4}
\usepackage{amsfonts,amssymb,amsmath,amsthm}
\usepackage{graphicx}

\begin{document}
\title{Symmetries in noncommutative field theories:
Hopf versus Lie\thanks{Based on a talk given at the Workshop
on Quantum Field Theory and Representation Theory,
 S\~ao Paulo, August 21-24, 2007.}}
\author{
D.~V.~Vassilevich\thanks{On leave from V.~A.~Fock Institute of
Physics, St.~Petersburg University, Russia.
E.mail:\ {\texttt{dmitry(at)dfn.if.usp.br}}}\\
{\it Instituto de F\'isica, Universidade de S\~ao Paulo,}\\ {\it
Caixa Postal 66318 CEP 05315-970, S\~ao Paulo, S.P., Brazil}}

\maketitle
\begin{abstract}
I discuss motivations for introducing Hopf algebra symmetries in
noncommutative field theories and briefly describe twisting of 
main symmetry transformations. New results include an extended list
of twisted gauge invariants (which may help to overcome the problem
of inconsistency of equations of motion) and a gauge-covariant
twist operator (leading to a gauge-covariant star product).
\end{abstract}
\section{Introduction}
The last decade has witnessed an ever growing interest to noncommutative
(NC) field theories \cite{Douglas:2001ba} motivated by a number of
physical applications. In such theories, the coordinates on the space-time
manifold $\mathcal{M}$ do not commute, meaning that the algebra of functions
on $\mathcal{M}$ is deformed to an associative but not commutative algebra.
This reminds us of a very interesting development in mathematics, namely
of the discovery of NC geometry \cite{ConnesBook}.

Symmetry is a guiding principle for constructing field theories.
General properties of symmetry transformations are discussed in 
sec.\ \ref{sec-sym}. Moyal-type noncommutativity is introduced in
sec.\ \ref{sec-NC}. These two sections are written for non-experts
and can be omitted by a more experienced reader.

In sec.\ \ref{sec-tro} we give an overview of several attempts
to deform the symmetries together with the algebra of functions
without deforming the Leibniz rule (which is the way in which
symmetry generators act on products).
We find that neither of this attempts was fully successful.
Either only a part of the symmetries can be preserved, or
the price to pay is the absence of a closed form of the action.

In sec.\ \ref{sec-twi} we introduce twisted symmetries which
eventually lead to a Hopf algebra structure of the symmetry
transformations. We then discuss features and problems of this
approach focussing on twisted gauge symmetries. We outline
solutions of some of the problems (which are new results first
reported in this article).
\subsection{Symmetries in field theory}\label{sec-sym}
To define a classical field theory one needs a smooth manifold
$\mathcal{M}$, called
the space-time. In this paper we shall consider a real plane 
$\mathbb{R}^n$ and a real torus $\mathbb{T}^n$ exclusively.
Fields are sufficiently smooth sections of a vector bundle
over $\mathcal{M}$.
The dynamics of classical field theory is governed by an
action $S$, which is a functional of fields. 
All quantities entering $S$ are subdivided into dynamical variables
and parameters. The difference between them is that one has to
vary the action with respect to the dynamical variables in
order to obtain equations of motion, while the parameters
are kept constant. To distinguish between dynamical and non-dynamical
variables one either uses some outside knowledge (for example,
one knows that electric charge of the proton is always the same,
so that it does not make sense to vary it in the action),
or checks whether resulting equations describe a meaningful
dynamics.

Let us consider a simple example. Let $\mathcal{M}$ is a two-dimensional
Minkowski space. Consider a scalar field $\phi$ (a section of trivial
line bundle) with the action
\begin{equation}
S=\int dx dt \left( (\partial_t\phi)^2 -(\partial_x\phi)^2 -m^2\phi^2
\right)
\end{equation}
which describes free propagation of a massive spinless particle.
Vanishing of the variation of $S$ with respect to $\phi$ yields a
wave equation $0=\delta S/\delta\phi = (-\partial_t^2 +\partial_x^2 -m^2)
\phi$. Variation of $S$ with respect to $m^2$ produces the condition
$\int dx\,dt\, \phi^2=0$ which has only a trivial solution $\phi \equiv 0$.
This confirms that mass of the particle must not be varied.

Symmetries are very important in physics. Symmetry is a transformation
of the dynamical variables which leaves the action invariant. Parameters
in the action are {\it not} transformed.

The most important global symmetry is the Poincare symmetry. The invariance
with respect to global translations is assured by the translational invariance 
of the space-time integration. The action is invariant with respect to
the rotations and Lorentz boosts if all vector indices are contracted
in pairs with the help of an invariant bi-linear form. By making translations
local (i.e. position-dependent) one obtains the diffemorphism transformations.
Any Poincare-invariant action can be made also diffeomorphism invariant
by introducing a (pseudo-) Riemannian metric to contract indices, and
corresponding connection and volume element.

Fundamental interactions of the Standard Model of elementary particles
correspond to gauge symmetries with the gauge group 
$SU(3)\times SU(2)\times U(1)$. All matter fields belong to some
unitary representations of this gauge group. Under infinitesimal
gauge transformations they transform as $\phi(x) \to \phi + \delta_\alpha
\phi$, $\delta_\alpha \phi =
\alpha (x)\phi (x)$,
where $\alpha (x)=\alpha_a(x) T^a$ with $T^a$ being the generators of
corresponding Lie algebra. Gauge fields $A_\mu$
then correspond to the connections. They have values in the Lie algebra
and transform according to the rule $\delta_\alpha A_\mu =-\partial_\mu
\alpha +[\alpha,A_\mu]$. It is easy to check that that 
${\rm tr} (F_{\mu\nu}F^{\mu\nu})$ with $F_{\mu\nu}=\partial_\mu A_\nu -
\partial_\nu A_\mu + [A_\mu,A_\nu]$ is gauge invariant.
\subsection{Noncommutative geometry and field theory}\label{sec-NC}
To describe an NC  deformation of a given manifold $\mathcal{M}$
one takes the algebra $\mathcal{A}$ of smooth functions on $\mathcal{M}$
and deforms is to an algebra $\mathcal{A}_\theta$ which is usually
assumed to be associative but not commutative. In the sense of Gelfand
and Naimark this algebra defines an NC manifold. Practically, one takes
the point-wise product $\mu : \mathcal{A}\otimes \mathcal{A}\to \mathcal{A}$,
$\mu (f_1 \otimes f_2)(x) = f_1(x)\cdot f_2(x)$ and replaces it with a
deformed product $\mu_\star$.

Let us now construct an NC version\footnote{A more refined construction
of the NC torus can be found in Ref. \cite{ConnesTm}.}
 of $\mathbb{T}^n$ and $\mathbb{R}^n$
with the Moyal product (also called the Weyl-Moyal of the Groenewald-Moyal
product).
Consider a  twist operator
\begin{equation}
\mathcal{F}=\exp \mathcal{P},\qquad \mathcal{P}=-\frac i2 \theta^{\mu\nu}
\partial_\mu \otimes \partial_\nu \label{twop}
\end{equation}
(which is indeed a twist, i.e. it satisfies the twist condition,
see eq.\ (\ref{twir}) below).
$\theta^{\mu\nu}$ is a skew-symmetric matrix (called an NC parameter).
The Moyal product $\mu_\star$ is then obtained by twisting the point-wise
product:
\begin{equation}
\phi_1\star \phi_2 \equiv \mu_\star (\phi_1 \otimes \phi_2)\equiv
\mu \circ \mathcal{F}^{-1} (\phi_1 \otimes \phi_2) \end{equation}
The algebra of smooth functions on $\mathbb{R}^n$ or $\mathbb{T}^n$
equipped with this star product is associative but not commutative.

To construct an NC counterpart of a usual (commutative) field theory
one takes an action and replaces all point-wise products by the
star-products. Of course, this prescription fails to give a unique
result since the expressions like $\phi_1 \star \phi_2 - \phi_2 \star \phi_1$
vanish in the commutative limit. It is natural to require that at least
the number of global and local symmetries is preserved by the deformation.
\section{Troubles with Lie algebra symmetries}\label{sec-tro}
The problem with symmetries in NC models is that the matrix $\theta^{\mu\nu}$
is a parameter rather than a dynamical variable. Therefore, $\theta^{\mu\nu}$
must not be transformed. However, $\theta^{\mu\nu}$ enters the twist operator
(\ref{twop}) is if it were a tensor. Precisely this inconsistency makes it
impossible to preserve usual Poincare and diffeomorphism invariances 
in NC theories.

Consider gauge transformations in the 
commutative case. Let us take two infinitesimal
gauge transformations with the parameters $\alpha (x)=\alpha_a (x)T^a$,
$\beta (x)=\beta_a (x) T^a$. Their commutator 
\begin{equation*}
[\alpha (x),\beta (x)]=[T^a,T^b] \alpha_a(x)\beta_b(x)
\end{equation*}
is again a gauge transformation.
In the NC case, a natural generalization of the gauge transformations
is $\delta_\alpha \phi = \alpha \star \phi$. 
The commutator of two consequent
transformations reads
\begin{eqnarray}
&&\alpha (x) \star \beta (x) -\beta (x) \star \alpha (x)=\nonumber\\
&&=\frac 12 [T^a,T^b] (\alpha_a \star \beta_b + \beta_b \star \alpha_a)+
\frac 12 \{ T^a,T^b\}  (\alpha_a \star \beta_b - \beta_b \star \alpha_a)
\nonumber
\end{eqnarray} 
This has to be a gauge transformation again.
Therefore, the set of generators $T^a$ must be closed with respect to both
commutators and anti-commutators. This requirement imposes very severe
restrictions on allowed gauge groups and their representations
\cite{Chaichian:2001mu}, which are not compatible with symmetries 
of the real World. In short, only the $U(n)$ type gauge generators
in fundamental or adjoint representation are allowed\footnote{
The use of semi-infinite Wilson lines allows to soften the
restrictions on representations \cite{Arai:2007dm}.}, though the
standard model of elementary particles requires more.

\subsection{The Seiberg-Witten map}
In 1999 Seiberg and Witten made an amazing discovery \cite{Seiberg:1999vs}.
They proposed a map which relates commutative and noncommutative gauge
theories. Due to this map, gauge symmetries of NC models may be
realized through standard commutative transformations of commutative
fields. In this way arbitrary gauge group can be realized. Although
the Seiberg-Witten map remains the main tool to study experimental
consequences of noncommutativity through modifications of the standard
model (see, e.g., \cite{Buric:2007qx} and references therein),
this cannot be considered as a complete solution of the problem.
The noncommutative fields are expressed as power series in $\theta$
with growing powers of commutative fields and derivatives (and just
a few terms are actually known explicitly). It is clear, that not
all effects can be studied in the framework of a $\theta$-expansion.
In quantum case, the models obtained through the Seiberg-Witten map
have (rather predictable) problems with renormalization, see
\cite{Wulkenhaar:2001sq}.
\subsection{Symplectic diffeomorphisms}
Since it does not look possible to make an NC theory invariant with respect
to all diffeomorphisms, it seems natural to consider a subalgebra
generated by the vector fields of the form
\begin{equation}
\xi^\mu(x) = \theta^{\mu\nu}\partial_\nu f(x),\label{sdif}
\end{equation}
which preserves $\theta^{\mu\nu}$ (under standard action of 
diffeomorphisms on a tensor) and try to construct a gravity theory
basing on such a symmetry \cite{Calmet:2005qm}. The transformations 
generated by the fields (\ref{sdif}) preserve the volume element.
Therefore, one deals with the so called unimodular gravity theories.
Although this approach gave rise to many interesting results over the
recent years, the group of symplectic diffeomerphisms is
too small to be the only symmetry of general relativity, even on
an NC manifold.

\subsection{Stability of NC Jackiw-Teitelboim model}
Another idea how to achieve a richer symmetry structure in NC theories
was suggested by two-dimensional dilaton gravity models 
\cite{Grumiller:2003df}. After a suitable field redefinition almost
all interesting models of that type can be written in the form
\begin{equation}
S_{\rm dil}=\int d^2x \varepsilon^{\mu\nu} (\phi \partial_\mu\omega_\nu
+\phi_a D_\mu e_\mu^a - \varepsilon^{ab}e_\mu^a e_\nu^b V(\phi)),
\label{2Ddil}
\end{equation}
where both kind of indices $a,b$ and $\mu,\nu$ take values $0,1$, $e_\mu^a$
is the zweibein, $\varepsilon^{\mu\nu}$ is an antisymmetric Levi-Civita
symbol ($\varepsilon^{10}=\varepsilon_{01}=1$). $D_\mu$ is a covariant 
derivative with the spin-connection $\omega_\mu \varepsilon^a_{\ b}$.
$\phi$ is a scalar field called the dilaton, and $\phi_a$ is an auxiliary
field (which essentially generates the torsion constraint). {\it Any} choice
of the potential $V(\phi)$ leads to a consistent model.

In the particular case of linear $V(\phi)$ one obtains the Jackiw-Teitelboim
model \cite{JT1,JT2,JT3}. This model is equivalent to a topological $SU(1,1)$
theory. The gauge group $SU(1,1)$ cannot be closed in the NC case, but
$U(1,1)$ can. By extending the model to an NC $U(1,1)$ topological theory
one arrives at the action \cite{Cacciatori:2002ib}
\begin{equation}
S^{(0)}=\frac{1}{4}\int d^{2}x\,\varepsilon^{\mu\nu}\left[\phi_{ab}\star
\left(R_
{\mu\nu}^{ab}-2\Lambda e_{\mu}^{a}\star e_{\nu}^{b}\right)
-2\phi_{a}\star T_{\mu
\nu}^{a}\right],
\label{actJT}
\end{equation}
where $R_{\mu\nu}^{ab}$ and $T_{\mu\nu}^{a}$ are noncommutative generalizations
of curvature and torsion which now depend on two connections, $\omega_\mu$ and
$b_\mu$. There is also a new dilaton field $\psi$ which enters the action
(\ref{actJT}) through the combination $\phi_{ab}:=\phi \varepsilon_{ab}
-i\eta_{ab}\psi$ with $\eta_{ab}={\rm diag}\, (+1,-1)_{ab}$. Together with
extending the gauge group one has to introduce new fields in the theory,
but this is a relatively moderate price to pay since these new fields decouple
in the commutative limit\footnote{A similar procedure can be also done
in higher dimensions (see, e.g, \cite{Chamseddine:2000zu}), but there
one has to add more fields, ant their decoupling in the commutative
limit is not automatic.}
and, hence, can be made invisible for present day
experiments. Besides, this model is surprisingly easy to quantize
\cite{Vassilevich:2004ym}.
One can find {\it all} quantum corrections by the methods
developed earlier in the commutative case \cite{Kummer:1996hy}.

It seems natural to look for a deformation of the action (\ref{actJT})
which, in the analogy with the commutative action (\ref{2Ddil}),
would have a proper number of local symmetries, though, maybe, with
a non-linear algebra, but still closed under the commutation and
with the standard Leibniz rule. The action (\ref{actJT})
is linear in the dilaton fields. To analyze the deformations one adds
to (\ref{actJT}) all possible quadratic terms without explicit derivatives
on the dilatons which are also real and preserve global Lorentz 
symmetry\footnote{In two dimensions 
$\theta^{\mu\nu}\simeq \varepsilon^{\mu\nu}$ is invariant under global Lorentz
transformations.}. For example, one can add $\varepsilon^{\mu\nu}
\varepsilon_{ab}e^a_\mu\star e^b_\nu \star \phi^2$ with an arbitrary
coefficient. There are seven independent terms \cite{Vassilevich:2006uv}.
In principle, one should also consider arbitrary deformations of local
symmetries and the solve the conditions that the action is invariant
under such symmetries. Fortunately, there is a short cut. One can use
the canonical formalism \cite{Vassilevich:2005fk} to check the closure
of the constraint algebra. The result is negative: no quadratic deformation
of (\ref{actJT}) preserves the number of local symmetries.

In two dimensions there is a similar result on the $\kappa$-Poincare algebra
which is quantum deformation of usual Poincare \cite{Lukierski:1991pn}.
Gravity theories in two dimensions with local $\kappa$-Poincare symmetry
but undeformed Leibniz rule are equivalent to undeformed theories
with local Poincare symmetry \cite{Grumiller:2003df}.
\section{Twisted symmetries}\label{sec-twi}
Since neither of the attempts to extend the standard (Lie algebra) approach
to symmetries to NC theories was completely successful, one arrives at an
idea to make a more substantial modification of the very concept of
symmetries. 

If we know the action of a symmetry generator $\alpha$ on the fields
$\phi_1$ and $\phi_2$ then the action of $\alpha$ on a tensor product
is defined by the so-called co-product $\Delta(\alpha)$.
In the case of a Lie algebra symmetry the coproduct is primitive,
$\Delta (\alpha)=\Delta_0(\alpha)=\alpha \otimes 1 + 1 \otimes \alpha$,
so that we have the usual Leibniz rule
\begin{equation}
\alpha (\phi_1 \otimes \phi_2) = \Delta_0(\alpha) (\phi_1 \otimes \phi_2)
=(\alpha \phi_1)\otimes \phi_2 + \phi_1 \otimes (\alpha \phi_2).
\label{prim}
\end{equation}

The primitive coproduct $\Delta_0$ is not the only possible coproduct.
To be able to discuss various coproducts systematically we need to
make a Hopf algebra out of our symmetry generators. Consider a Lie
algebra $G$ and its universal enveloping algebra $H=\mathcal{U}(G)$.
Then $H$ is an associative unital algebra. A coproduct is an
algebra homomorphism $\Delta : H\to H\otimes H$ which satisfies the
coassociativity relation
\begin{equation}
(\Delta \otimes 1)\circ \Delta = (1\otimes \Delta )\circ \Delta.
\label{coass}
\end{equation}
To complete the Hopf algebra structure one also has to introduce
an antipode and a counit, both satisfying certain relations with
the coproduct and between themselves. We shall not need these elements.
The interested reader can consult any textbook on Hopf algebras
or on quantum groups. A nice simple introduction can be found in
\cite{Szabo:2006wx}.

Suppose we have a twist element $\mathcal{F}\in H\otimes H$
satisfying the relation \cite{Reshetikhin:1990ep}
\begin{equation}
(\mathcal{F}\otimes 1)(\Delta \otimes 1)\mathcal{F}=
(1\otimes \mathcal{F})(1\otimes \Delta)\mathcal{F}
\label{twir}
\end{equation}
(and another relation involving counit). Then we may define another
(twisted) coproduct
\begin{equation}
\Delta_{\mathcal{F}}=\mathcal{F}\Delta \mathcal{F}^{-1}\label{twic}
\end{equation}
(also twisting the counit and the antipode). Suppose now that our algebra
$G$ contains space-time translations which are represented by partial
derivatives. Then the twist element defined in eq.\ (\ref{twop})
above belongs to $H\otimes H$. One can check that
the equation (\ref{twic}) is satisfied by (\ref{twop}) for $\Delta =\Delta_0$.

The action of a generator $\alpha$ on the star-product of fields is defined
as follows
\begin{equation}
\alpha (\phi_1\star\phi_2)=\mu_\star (\Delta_{\mathcal{F}}(\alpha) 
\phi_1\otimes\phi_2)=\mu\circ \mathcal{F}^{-1} (\Delta_{\mathcal{F}}(\alpha) 
\phi_1\otimes\phi_2)
\label{alstar}
\end{equation}
Here and everywhere below $\Delta_{\mathcal F}\equiv (\Delta_0)_{\mathcal F}
={\mathcal{F}}\Delta_0{\mathcal{F}}^{-1}$. In a sense, twisting pushes
the generator $\alpha$ through the star-product, so that the star-product
itself is not transformed. Therefore, it becomes much easier to construct
invariants.

The idea to twist physical symmetries appeared
already in \cite{KM}, though with a different twist. The same twist
as above but without analysing invariants was suggested first
in \cite{Oeckl:2000eg}. The real break through came later, when
twisted Poincare symmetry of noncommutative field theories was constructed
\cite{Chaichian:2004za,Wess:2003da,Chaichian:2004yh}. Afterwards twisted
conformal symmetries \cite{Matlock:2005zn,Lizzi:2006xi}, twisted
diffeomorphisms \cite{Aschieri:2005yw,Aschieri:2005zs}, and twisted
gauge symmetries \cite{Vassilevich:2006tc,Aschieri:2006ye} were constructed
(to mention bosonic symmetries only). Let us consider in some detail
twisted gauge symmetries \cite{Vassilevich:2006tc,Aschieri:2006ye}. This
particular case is chosen since (i) it is rather simple, (ii) the topic
is still quite controversial, and (iii) I have something new to say
on this subject.

To be concrete, let us consider a theory describing some scalar field
$\phi$ and gauge fields (connections) $A_\mu$. Gauge transformations
of these fields with the parameter $\alpha(x)$ can be written as
\begin{equation}
\alpha : \Phi \to \Phi + \delta_\alpha \Phi,\qquad
\delta_\alpha \Phi=R(\alpha) \Phi ,\label{aPh}
\end{equation}
where
\begin{equation}
\Phi \equiv \left( \begin{array}{c} \phi \\ A_\mu \\ 1 \end{array}\right),
\qquad
R(\alpha)=\left( \begin{array}{ccc} \alpha (x) & 0 & 0 \\
0 & {\rm ad}(\alpha) & -\partial_\mu \alpha \\
0 & 0 & 0 \end{array} \right).\label{Ral}
\end{equation}
The twisted coproduct reads in the $\theta$-expansion
\begin{eqnarray}
\Delta_{\mathcal{F}}(\alpha)&=&R(\alpha) \otimes 1 + 1 \otimes R(\alpha)
-\frac i2 \theta^{\mu\nu} ([\partial_\mu,R(\alpha)\otimes \partial_\nu +
\partial_\mu \otimes [\partial_\nu,R(\alpha)])\nonumber\\
& &-\frac 18 \theta^{\mu\nu}\theta^{\rho\lambda} ([\partial_\mu,
[\partial_\rho,R(\alpha)]]\otimes \partial_\nu \partial_\lambda
+\partial_\mu\partial_\rho \otimes [\partial_\nu,[\partial_\lambda,R(\alpha)]]
+\mathcal{O}(\theta^3).\nonumber
\end{eqnarray}
Now one can start constructing invariants. One immediately finds a lot
of invariants involving the scalar field and a very important invariant
of the gauge field
\begin{equation}
{\rm tr}(F_{\mu\nu}\star F^{\mu\nu}) \label{FF}
\end{equation}
constructed from an NC generalization of the field strength (bundle
curvature)
\begin{equation}
F_{\mu\nu}=\partial_\mu A_\nu - \partial_\nu A_\mu + A_\mu \star A_\nu
-A_\nu \star A_\mu \,.\label{AF}
\end{equation}

The main advantage of this scheme is that any gauge group can be realized,
but there are also drawbacks.  One of them 
was noted already in \cite{Aschieri:2006ye}
(for a more elaborate discussion see \cite{Giller:2007gq}). It was
demonstrated that the action with the density (\ref{FF}) leads to
inconsistent equations of motion unless one adds more vector fields
with the values in the enveloping algebra of the original gauge
algebra. However, this statement refers to just {\it one} possible 
deformation (\ref{FF}) of the commutative action for gauge fields.
There are others. The key observation which has led to twisted gauge 
invariance of (\ref{FF}) is that $F_{\mu\nu}$ is twisted gauge
covariant,
\begin{equation}
\delta F_{\mu\nu}(x) = \alpha_a(x)\cdot [T^a,F_{\mu\nu}(x)]\,.
\label{tgc}
\end{equation}
The action of symmetry transformations in (\ref{tgc}) is reducible.
The components of $F_{\mu\nu}$ which belong to the Lie algebra transform
through themselves. The same is true for the components which are proportional
to anti-commutators $\{ T^a,T^b \}$. Therefore, instead of one covariant
object we have two:
\begin{eqnarray}
&&F_{\mu\nu}^{(1)}
=\partial_\mu A_\nu - \partial_\nu A_\mu + (A_{a\mu} \star A_{b\nu}
+A_{b\nu} \star A_{a\mu})\frac 12 [T^a,T^b],\label{F1} \\
&&F_{\mu\nu}^{(2)}=(A_{a\mu} \star A_{b\nu}
-A_{b\nu} \star A_{a\mu})\frac 12 \{ T^a,T^b\},\label{F2}
\end{eqnarray}
and instead of a single invariant (\ref{FF}) we have a two-parameter
family 
\begin{equation}
{\rm tr}( F^{(1)}_{\mu\nu}\star F^{(1)}_{\mu\nu} +
g_1 F^{(2)}_{\mu\nu}\star F^{(2)}_{\mu\nu} +
g_2 F^{(1)}_{\mu\nu}\star F^{(2)}_{\mu\nu}) \,.\label{famin}
\end{equation}
It is quite possible (anyway, not excluded by the results of
\cite{Aschieri:2006ye,Giller:2007gq}) that for some choice of the
parameters $g_1$ and $g_2$ and for some gauge groups the equations
of motion become consistent.

Another criticism of the scheme based on twisted gauge transformations
appeared in \cite{Chaichian:2006we}. The authors of \cite{Chaichian:2006we}
claimed that this scheme contradicts the gauge principle since it
implicitly assumes that the action of the Poincare generators 
appearing in the twist does not
change representation of the gauge group. In other words, the twist
operator is not covariant.
According to a more moderate point of view
 \cite{Wess:2006cm} it is enough to have proper commutation relation 
between the symmetry generators and a coassociative coproduct, so that
one has a Hopf algebra based on a proper Lie algebra. Gauge covariance
of twist operator is not required. Regardless of whether covariance of
the twist must or must not be included in a proper formulation of the
gauge principle\footnote{This question may only be answered by studying
physical consequences of both schemes.} it is interesting on its own
right to find a gauge covariant twist operator. Simply replacing
partial derivatives by covariant derivatives does not work since
the resulting star-product is non associative \cite{Chaichian:2006wt}
(see also \cite{Harikumar:2006xf} for a related discussion in curved
space).

The main ingredient of the construction below is a {\it trivial} 
connection\footnote{Trivial connections were used in star products
already in \cite{Batalin:1990yk} though without any relation to
gauge symmetries.}
$\tilde A_\mu = U (\partial_\mu U^{-1})$, where $U(x)$ is an element of
a finite-dimensional Lie group (which will later become the gauge
group of some model). Obviously, the covariant derivatives
$\tilde \nabla_\mu = \partial_\mu +\tilde A_\mu$ commute, and,
as a consequence, the operator
\begin{equation}
\mathcal{F}_U=\exp \mathcal{P}_U,\qquad \mathcal{P}_U
=-\frac i2 \theta^{\mu\nu}
\tilde \nabla_\mu \otimes \tilde\nabla_\nu \label{FU}
\end{equation}
satisfies the twist condition and can be used to construct an
associative star product
\begin{equation}
\phi_1 \star_U\phi_2
=\mu \circ \mathcal{F}_U^{-1} (\phi_1\otimes \phi_2).
\label{stU}
\end{equation}
More explicitly, if $\phi_1$ and $\phi_2$ transform according to
representations $R_1$ and $R_2$ of the gauge group respectively,
\begin{equation}
\phi_1 \star_U\phi_2 = (R_1\otimes R_2)(U)\cdot (R_1(U^{-1})\phi_1
\star R_2(U^{-1})\phi_2),\label{stRR}
\end{equation}
where $\star=\star_1$ is the Moyal product. The field $A_\mu$ does not 
belong to a linear representation of the gauge group. To construct
a $\star_U$ product involving $A_\mu$ one has to exponentiate
(\ref{Ral})
\begin{equation}
\Phi \equiv \left( \begin{array}{c} \phi_1 \\ A_\mu \\ 1 \end{array}\right),
\qquad
R(U)=\left( \begin{array}{ccc} R_1(U) & 0 & 0 \\
0 & {\rm Ad}(U) & U \partial_\mu U^{-1} \\
0 & 0 & 1 \end{array} \right).\label{RU}
\end{equation}
The rest is straightforward, in particular we obtain
\begin{equation}
A_\mu \star_U \phi_1 = U( (U^{-1}\partial_\mu U +U^{-1}A_\mu U)\star (U^{-1}
\phi_1)) + U\partial_\mu U^{-1}\cdot \phi_1 \label{Amup}
\end{equation}
(both $U$ and $A_\mu$ are taken in the representation $R_1$). 

One can now proceed with constructing twisted gauge invariants in parallel
to what we have outlined above in the case of usual Moyal product.
In particular, one finds the property (\ref{tgc}) for the field strength
(\ref{AF}) with $\star$ replaced by $\star_U$. The new invariants constructed
with $\star_U$ have a very important property which old invariants
(constructed with $\star$) do not have: they become \emph{true gauge 
invariants with undeformed Leibniz rule} if the transformation rules
(\ref{aPh}) for $\phi$ and $A_\mu$ are accompanied by a transformation
of $U$, $\delta_\alpha U=\alpha \cdot U$. Therefore, we have just constructed
an NC gauge theory with standard gauge transformations and an arbitrary
gauge group. This makes twisting of gauge transformations unnecessary.

Now we have to return to the beginning of this article. Since $U$
is transformed under some symmetry transformations, it has to be a dynamical
field. There are two distinct options.

(i) $U$ is an independent dynamical field. In this case one has to
add a suitable action for $U$ to make corresponding equations of
motion elliptic (or hyperbolic in the Minkowski space).

(ii) $U$ is a an already existing degree of freedom. One can, for
example, identify $U$ with longitudinal degrees of freedom of $A_\mu$
by using the representation $A_\mu=UA_\mu^{T} U^{-1}+U\partial_\mu U$,
where $A^{T}$ satisfies some gauge condition. E.g., the Lorenz condition
$\partial^\mu A_\mu^{T}=0$ can be taken. Such a scheme depends,
however, on gauge conditions imposed on $A_\mu^T$.

At present, it is not clear which of these two options (if any) is
compatible with physics.
\section{Conclusions}
The main message of this paper is that usual concept of symmetries with the
standard Leibniz rule seems to be insufficient for noncommutative
field theories. The Leibniz rule (coproduct) must be deformed (twisted).
Therefore, Hopf algebras arise in the field theory context. Over the last
three years twisted counterparts were defined for
almost all important physical symmetries. This subject is still a rather
new one. Many problems remain open. We outlined possible solutions for
two of such problems in the case of gauge symmetries.

We considered classical field theories only. An appropriate quantization
scheme respecting the twisted symmetries is currently under debate,
see recent papers \cite{Tureanu:2006pb,Zahn:2006wt,Fiore:2007vg,
Balachandran:2007kv,Riccardi:2007bj} 
and references therein.

\subsection*{Acknowledgement}
I am grateful to A.~Pinzul for interesting discussions.
This work was supported in part by FAPESP.

\end{document}